\documentclass[usenatbib]{mn2e}
\usepackage{graphicx}
\usepackage{ifthen}

\def\ltsima{$\; \buildrel < \over \sim \;$}
\def\simlt{\lower.5ex\hbox{\ltsima}}
\def\gtsima{$\; \buildrel > \over \sim \;$}
\def\simgt{\lower.5ex\hbox{\gtsima}}
\def\kms{{\rm\,km\,s^{-1}}}

\def\kpc{{\rm\,kpc}}

\def\msun{{\rm\,M_\odot}}

\def\pc{{\rm\,pc}}

\newcommand{\fmmm}[1]{\mbox{$#1$}}
\newcommand{\scnd}{\mbox{\fmmm{''}\hskip-0.3em .}}

\def\AA{$\; \buildrel \circ \over {\rm A}$}

\def\UseFigs{1}

\def\degg{\hbox{$\null^\circ$\hskip-3pt .}}

\def\s{\ifmmode \widetilde \else \~\fi}
\def\={\overline}

\def\spose#1{\hbox to 0pt{#1\hss}}

\def\lta{\mathrel{\spose{\lower 3pt\hbox{$\mathchar"218$}}
     \raise 2.0pt\hbox{$\mathchar"13C$}}}
\def\gta{\mathrel{\spose{\lower 3pt\hbox{$\mathchar"218$}}
     \raise 2.0pt\hbox{$\mathchar"13E$}}}
\def\Dt{\spose{\raise 1.5ex\hbox{\hskip3pt$\mathchar"201$}}}    
\def\dt{\spose{\raise 1.0ex\hbox{\hskip2pt$\mathchar"201$}}}    

\def\dotsfill{\leaders\hbox to 1em{\hss.\hss}\hfill}

\def\Gyr{{\rm\,Gyr}}

\loadboldmathitalic 
\title[The course of the Andromeda stream]
{Taking measure of the Andromeda halo: \\
a kinematic analysis of the  giant stream surrounding M31}
\author[R. Ibata, S. Chapman, A. M. N. Ferguson, M. Irwin, G. Lewis,\\ A. McConnachie]
{R. Ibata$^{1}$, S. Chapman$^{2}$, A. M. N. Ferguson$^{3}$, M. Irwin$^{4}$, 
G. Lewis$^{5}$, A. McConnachie$^{4}$\\
$^{1}$
Observatoire de Strasbourg, 11, rue de l'Universit\'e, F-67000, Strasbourg, 
France\\
$^{2}$
California Institute of Technology, Pasadena, CA 91125, U.S.A\\
$^{3}$
Max-Plank Institut f\"ur Astrophysik, Karl-Schwarzschild-Str. 1, Postfach
1317, D-85741, Garching, Germany\\
$^{4}$
Institute of Astronomy, Madingley Road, Cambridge, CB3 0HA, U.K.\\
$^{5}$
Institute of Astronomy, School of Physics, A29, University of Sydney, NSW
2006, Australia}

\date{\today}
\begin{document} 
\maketitle 
\begin{abstract}
We present a spectroscopic survey of the giant stellar stream found in
the halo of  the Andromeda galaxy. Taken with  the DEIMOS multi-object
spectrograph  on the  Keck2  telescope, these  data  display a  narrow
velocity  dispersion of  $11\pm3\kms$, with  a steady  radial velocity
gradient of $245 \kms$ over the $125 \kpc$ radial extent of the stream
studied so  far.  This implies  that the Andromeda galaxy  possesses a
substantial  dark matter  halo.  We  fit the  orbit of  the  stream in
different galaxy potential models. In  a simple model with a composite
bulge, disk and  halo, where the halo follows  a ``universal'' profile
that is  compressed by  the formation of  the baryonic  components, we
find that  the kinematics  of the stream  require a total  mass inside
$125\kpc$ of  $M_{125} =  7.5^{+2.5}_{-1.3} \times 10^{11}  \msun$, or
$M_{125} >  5.4 \times  10^{11} \msun$ at  the 99\%  confidence level.
This is the first galaxy in  which it has been possible to measure the
halo  mass distribution  by such  direct dynamical  means over  such a
large distance range. The resulting orbit shows that if M32 or NGC~205
are  connected with  the stream,  they must  either trail  or  lag the
densest  region of  the stream  by more  than  $100\kpc$. Furthermore,
according to the best-fit orbit,  the stream passes very close to M31,
causing  its demise as  a coherent  structure and  producing a  fan of
stars that will  pollute the inner halo, thereby  confusing efforts to
measure  the properties of  genuine halo  populations.  Our  data show
that several  recently identified  planetary nebulae, which  have been
proposed as evidence for the existence  of a new companion of M31, are
likely members of the Andromeda Stream.
\end{abstract}

\begin{figure*}
\ifthenelse{\UseFigs=1}{
\includegraphics[angle=270,width=\hsize]{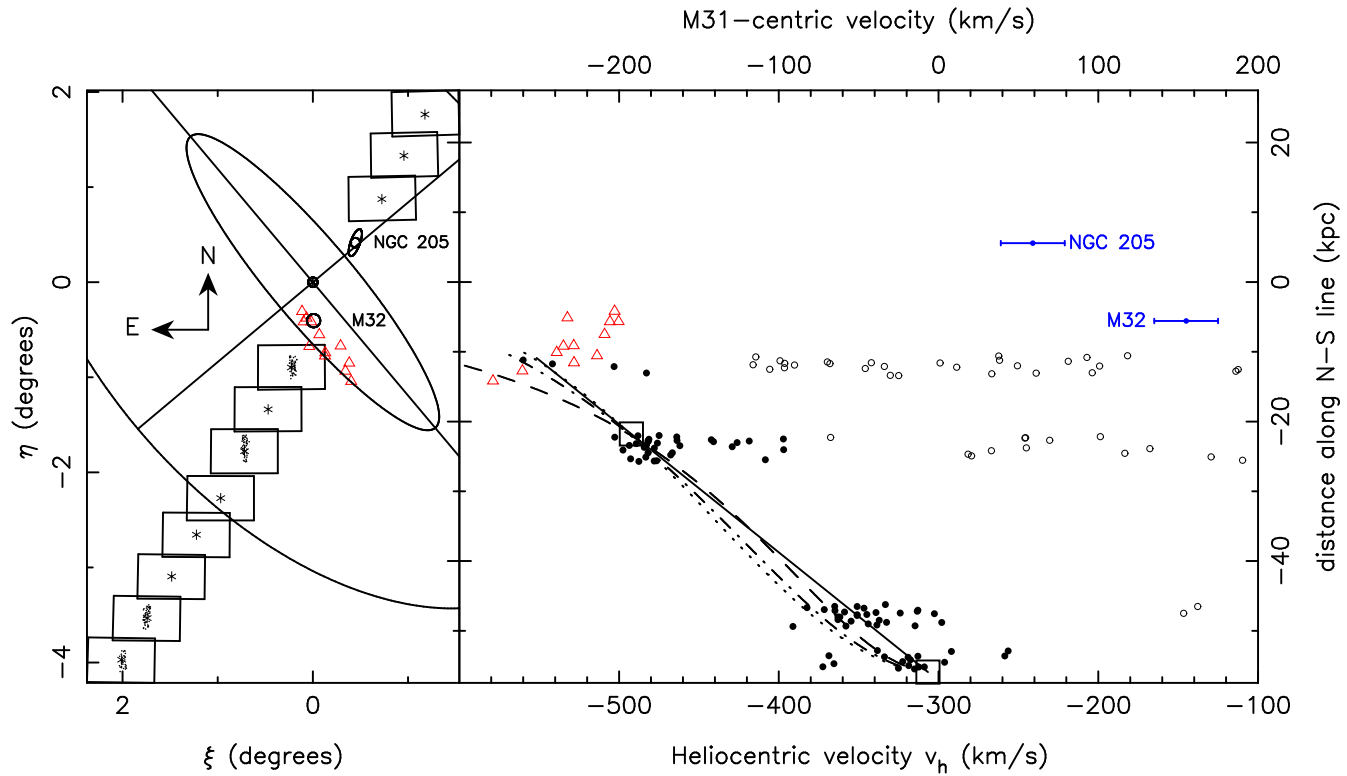}
}{}
\caption{The  left hand  panel  shows the  locations  of the  observed
fields on the plane of  the sky, in standard coordinates $(\xi,\eta)$.
Asterisk symbols denote field centres  of, from South to North, Fields
1--8 and 12--14 of our photometric survey of the Andromeda Stream with
the CFHT12K camera; while the large rectangles display the size of the
CFHT12K camera pointings. The Keck2  DEIMOS targets are shown as small
dots in the  centre of Fields 1, 2, 6 and  8.  Triangles represent the
positions  of the  planetary nebulae  identified  by \citet{morrison}.
The inner ellipse demarcates the approximate limit of the visible disk
of M31 at $2^\circ$ ($=27\kpc$  radius), and the outer ellipse shows a
segment  of   a  $50\kpc$  radius  ellipse   flattened  to  $c/a=0.6$,
corresponding to the  approximate limit of our INT  survey.  The right
hand panel  presents the radial velocity  measurements (full circles),
as a  function of  distance from M31  along the  declination direction
[$=780\tan(\eta)\kpc$].   The positions  of M32  and NGC~205  are also
shown. The  black line shows a straight-line  fit [$v_h(\eta)= -4244.8
\tan(\eta) - 610.9  \kms$] to the high negative  velocity edge of this
diagram.  Stars that  are more than $100\kms$ away  from this line are
marked  with open  circles.  The  two  large squares  show the  chosen
points  for  our  ``back-of-the-envelope''  calculation,  detailed  in
Section~3.  The  dashed, dot-dashed and dotted lines  show the orbital
paths in simple Kepler,  logarithmic and NFW potentials, respectively.
The velocities of the  \citet{morrison} PNe are represented again with
triangles.}
\end{figure*}

\section{Introduction}

Stellar streams represent the  visible remnants of the merging process
by which the halos of galaxies are built up. By studying these streams
we  can attempt  to unravel  the formation  of galactic  halos, seeing
when, how  and how many  small galaxies arrived and  were incorporated
into large galaxies \citep{helmi99,  johnston01}.  Streams are also of
great interest as  probes of the large-scale mass  distribution of the
dark halos they reside  in \citep{johnston99, ibata01a, zhao99}.  This
utility  stems from  the fact  that streams  from  low-mass disrupting
stellar systems trace the orbit of their progenitor, giving a means to
constrain the tangential motion of the stars in the stream.  The stars
must  move along  the  stream,  and the  magnitude  of the  tangential
velocity  must be such  that, when  the star  is integrated  along the
orbit,  it ends  up  with the  same  velocity as  the  stars that  are
currently downstream on the same orbit.

Giant stellar  streams may well  be common structures  around galaxies
\citep{pohlen, malin}.   Indeed, the  most conspicuous feature  in the
halo of the Milky Way is the giant rosette stream originating from the
Sagittarius  dwarf galaxy,  which contains  approximately half  of the
high  latitude ($|b|>30^\circ$)  intermediate age  stars  at distances
greater than  $15\kpc$ \citep{ibata01a, ibata02,  majewski}.  It would
appear that  the Milky Way has  not incorporated into the  Halo a more
massive galaxy than the Sagittarius dwarf over the last $\sim 7\Gyr$.

This naturally  leads to  the question of  whether the MW  has unusual
feeding  habits.    To  answer  this,  we  have   undertaken  a  large
photometric study of M31, using the  wide field cameras at the INT and
CFHT telescopes to  resolve stars over the entire  disk and inner halo
of  that  galaxy  \citep{ibata01b,ferguson02,mcconnachie}.   This  has
given us an unprecedented panoramic  view of the large scale and small
scale structure of a disk galaxy. The analysis of this huge dataset is
still in progress, but it  has already yielded some surprising results
regarding the incidence of substructure  in the halo of M31.  The most
prominent  of these  substructures  is a  stream-like over-density  of
stars near  the minor axis of  M31 \citep{ibata01b}, at  first sight a
facsimile of  the Sagittarius  Stream around the  Milky Way.   The red
giant branch (RGB) stellar density in the halo increases on average by
a  factor  of  two  in  the on-stream  regions  and  is  statistically
significant at the 50-sigma  level.  Interestingly, this stream points
toward the  Andromeda satellites M32  and NGC205, and is  aligned with
the outer  isophotes of NGC205, suggesting a  relationship between the
Andromeda Stream and these two dwarf galaxies.  If this interpretation
is correct, the  stream has to be the  result of previous interactions
with M31, as there is  otherwise not enough time to spatially separate
it from either of the two dwarf galaxies.

The proximity  of M31 provides us  with an opportunity  to undertake a
spectroscopic  survey of  individual stars  within the  stream; indeed
Andromeda offers the  only extragalactic giant galaxy in  which such a
study  can be  undertaken  with current  instrumentation,  as in  more
distant systems such halo substructure  would be smeared into very low
surface brightness features.

\begin{figure*}
\ifthenelse{\UseFigs=1}{
\includegraphics[angle=270,width=\hsize]{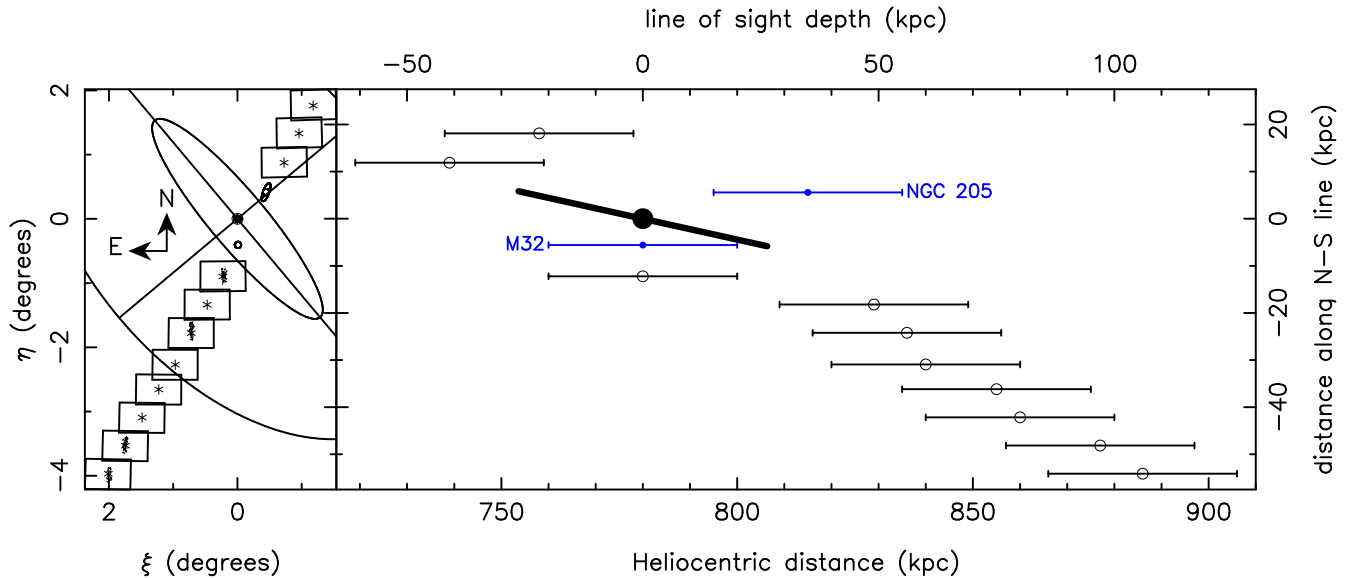}}{}
\caption{For  convenience, the  left-hand panel  reproduces  the chart
shown previously  in Figure~1. The  right-hand panel shows  a sideways
view of  the stream, drawn to  the same scale as  the left-hand panel,
which displays  the line of sight  depth of the  fields, together with
the positions of M32 and NGC~205.   The disk of M31 is highly inclined
to  our line  of sight  ($12\degg5$); the  thick line  is  a schematic
representation   of   a   disk   of  radius   $27\kpc$   inclined   at
$12\degg5$.  Evidently the  stream orbits  close to  the plane  of the
Andromeda galaxy.}
\end{figure*}

One useful  property of the  stream is that  it is on a  highly radial
orbit. It passes  very close to the centre of  M31, where a comparison
of the RGB tip of the stream with that of M31 itself shows them to lie
at  the same  distance  \citep{mcconnachie}; whereas  in the  furthest
field that it has been detected  to date, the peak of its giant branch
is 0.27~mag  fainter, indicating that it is  $106\pm20\kpc$ behind M31
\citep{mcconnachie}.  This  fortuitous alignment close to  the line of
sight, allows  us to measure  directly the potential gradient  over $>
100\kpc$, and hence measure the halo mass. 

The  layout  of this  paper  is  as  follows: section~2  presents  the
spectroscopic survey of the Andromeda  Stream, the results of which are
used in  section~3 to constrain  the mass of  the dark matter  halo of
M31,  and finally  in section~4  we  discuss the  significance of  the
results  and present the  conclusions of  this study.  Throughout this
work, we  assume a distance of  $780\kpc$ to M31  \citep{stanek} and a
systemic radial velocity of $-300\kms$ \citep{devaucouleurs}.

\section{The Andromeda halo Survey}

To  understand  the  nature  of  the  substructures  detected  in  our
panoramic halo surveys  of M31, we undertook a  follow-up programme to
measure  the  kinematics  of  individual  RGB stars  with  the  DEIMOS
spectrograph on the  Keck2 10m telescope.  On Sep  29-30 2002, we used
the  instrument in  the high-resolution  configuration with  the ${\rm
1200 l/mm}$ grating (giving access  to the spectral region 6400\AA\ to
9000\AA) to  obtain high quality spectra  of 768 stars in  9 fields in
M31. This  has given an  order of magnitude improvement  in statistics
over   previous   radial   velocity   surveys  of   the   M31   system
\cite{reitzel}.   The  present contribution  examines  data from  four
fields along the giant stellar stream.

Red giant branch stars were  selected for observation from our CFHT12K
camera survey, which covers the  region of the sky shown schematically
on the  left-hand panel of  Figure~1. Stars were selected  by choosing
point-sources with I-band  magnitudes between ${\rm 20.5 <  I < 22.0}$
and colours ${\rm  1.0 < V-I < 4.0}$.   Both metal-poor and metal-rich
populations  will  be  present  in  this selection  region  (see  e.g.
\citealt{mcconnachie}). The  number of target stars  per $16.9' \times
5'$ DEIMOS field is approximately 100.  The slitlet widths were milled
at $0\scnd7$ to match the  median seeing, and the slitlet-lengths were
always larger  than $5''$, giving access  to a good  estimation of the
local sky.

The  spectroscopic  images  were  processed  and  combined  using  the
pipeline software developed by  the ``DEEP'' consortium. This software
debiasses, performs a  flat-field, extracts, wavelength-calibrates and
sky-subtracts the spectra. After  extracting the spectra with a boxcar
algorithm,  the radial  velocities  of the  stars  were measured  with
respect  to spectra of  standard stars  observed during  the observing
run.  By  fitting the  peak of the  cross-correlation function  with a
Gaussian, an estimate of the radial velocity accuracy was obtained for
each  radial velocity  measurement.  The  accuracy of  these  data are
astonishing  for  such  faint  stars, with  typical  uncertainties  of
$5\kms$  to  $10\kms$.   Finally,  M31-centred radial  velocities  are
calculated by adding $300\kms$ to the Heliocentric values.

In the  4 stream fields 184  stars were measured  with radial velocity
uncertainties  less than  $25\kms$. The  right-hand panel  of Figure~1
shows  the  subset  of  125  stars  that  have  heliocentric  velocity
$v_{h}<-100\kms$, as a function of projected distance $d_\delta (= 780
\tan[\eta] \kpc)$ along the declination direction.  For clarity, stars
with $v_{h}>-100\kms$, which are primarily Galactic, have been omitted
from this diagram.  These four  fields correspond to CFHT Fields 1, 2,
4 and  8, and their locations on  the sky are marked  in the left-hand
panel. The Heliocentric distances of  the fields, as measured from the
magnitude  of   the  Tip   of  the  RGB   of  the   Stream  population
\citep{mcconnachie},  are   displayed  in  the   right-hand  panel  of
Figure~2.  The outermost field lies at ($\xi=-2\degg0, \eta=-4\degg1$)
at a  Heliocentric distance of $886  \pm 20\kpc$, that  is, it extends
out to  $125 \pm 17\kpc$  from the centre  of M31.  This is  the first
time that it has been possible  to measure the kinematics of a stellar
stream  over  such  a  huge  range  in  galactocentric  distance.   An
immediately striking  aspect of  these data is  the sharp edge  to the
distribution  at  negative velocities  in  Figure~1,  with a  velocity
gradient  that appears  to be  almost a  straight line.   The velocity
distribution  near  this  edge  is  very narrow,  as  demonstrated  by
Figure~3, where we show the  distribution of velocity offsets from the
straight line in  Figure~1.  We compare the data  between $-40\kms$ to
$40\kms$ in Figure~3 to a  Gaussian model using the maximum likelihood
technique,  and find  that the  dispersion is  $11 \pm  3  \kms$.  The
distribution appears slightly  skewed to positive velocities; defining
skewness   to   be   ${{1}\over{N}}   \sum_{j=1}^N   \Big[   {{x_j   -
\overline{x}}\over{\sigma}}  \Big]^3$  \citep[see e.g.,][]{press},  we
find that  the distribution of 82  stars in Figure~3 that  have $|v| <
100\kms$  have a  skewness of  $0.51$.  For  comparison,  the standard
deviation of the skewness of samples of 82 stars drawn from a Gaussian
distribution is  $0.43$, so the  observed sample is  not significantly
skewed. Some of the stars in  the skewed tail may be contaminants from
the halo of M31, and others may be stream stars more distant along the
line of  sight. The  nature of these  objects may become  clearer when
N-body simulations are fit to the stream.

\begin{figure}
\ifthenelse{\UseFigs=1}{
\includegraphics[angle=270,width=\hsize]{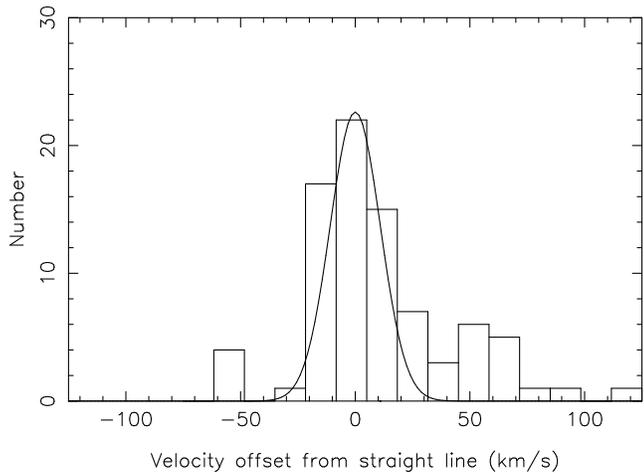}
}{}
\caption{The  distribution  of  velocities  about  the  straight  line
displayed  on the  right-hand  panel  of Figure~1.  The  stream has  a
narrowly-peaked velocity distribution at the position of that straight
line, with dispersion $11\pm3\kms$.}
\end{figure}

\section{Constraints on the dark matter halo}

Before fitting an orbit model  in a realistic galaxy potential through
the position and velocity data,  it is worth investigating what can be
learned from  a simple ``back-of-the  envelope'' analytic calculation.
We  assume that the  global potential  $\Psi$ at  the distance  of the
stream is approximated  by a spherical potential, and  that the stream
follows the  orbit of the  centre of mass  so all stars have  the same
total  energy   $E=  \Psi(r)  +   {{1}\over{2}}  \vec{v}(r)^2$,  where
$\vec{v}$  is the  three  dimensional  velocity of  a  star at  radial
position $r$.  If we assume further that the orbit is radial, which is
a reasonable approximation given the distance information in Figure~2,
then $\vec{v}(r) = v(\beta) / \cos(\gamma)$, where $v$ is the observed
(one-dimensional) radial  velocity, as a function  of angular distance
$\beta           =           \cos^{-1}[\cos(\xi)\cos(\eta)]          =
\sin^{-1}[{{r}\over{780\kpc}}  \sin(\gamma)]$  along  the stream,  and
$\gamma$  is  the  projection  angle  onto the  line  of  sight.   The
M31-Sun-Field~1       triangle       has       an       angle       of
$\cos^{-1}[\cos(-2\degg0)\cos(-4\degg1)]$,   and   lengths   $780\kpc$
(M31-Sun),  $886  \pm  20\kpc$  (Sun-Field~1), and  $125  \pm  17\kpc$
(Field~1-M31).  Thus the cosine of  the projection angle onto the line
of  sight  is $\cos(\gamma)=0.868\pm0.040$.   We  take  2 data  points
$(\xi_1 = 2\degg0, \eta_1 = -4\degg1)$ and $(\xi_2 = 0\degg7, \eta_2 =
-1\degg6)$, for which  the de-projected Andromeda-centric distance and
de-projected velocity  are: $(r_1 = 125 \pm  17 \kpc, v_1 =  -9 \pm 13
\kms)$, $(r_2 =  r_1\sin(\eta_2)/\sin(\eta_1) = 49 \pm 18  \kpc, v_2 =
-222 \pm 13 \kms)$, to  represent the stream. The uncertainty on $r_1$
is calculated  from the  projection and distance  uncertainties, while
the  uncertainties   on  $v_1$  and  $v_2$  are   estimated  from  the
$11\pm3\kms$  dispersion  combined  with the  projection  uncertainty.
These two points  correspond to the extremities of Fields  1 and 6 (we
do not choose Field 8, as this region is much deeper in the potential,
where  the disk  contribution is  significant).  In  calculating these
de-projected velocities, we have  assumed that the tangential velocity
of  M31 is  zero.  It  has long  been suspected  that  this tangential
velocity is  small given that  there is no  other large galaxy  in the
Local  Group to  provide significant  torque to  the Milky  Way  - M31
system \citep{kahn}.   \citet{DLB77} have  also pointed out  that this
tangential motion must be small, as there is otherwise not enough time
for M31  and the  Milky Way  to have almost  completed an  orbit about
each other in  the age  of the  Universe, as is  required by  the Local
Group timing  arguement.  The result of  \citet{einasto} is consistent
with this possibility; they find  that the transverse motion of M31 is
$60\pm30\kms$, under  the assumption that  M31 and the Milky  Way have
equal and opposite angular momenta.   To provide a crude assessment of
the effect of the possible  transverse motion of M31 on the parameters
derived from our  dataset, we also consider in  the analysis below the
consequence of a transverse motion of amplitude $300\kms$ (i.e.  equal
to the  radial velocity  component) in the  direction parallel  to the
Stream.

For  a Kepler  potential, the  two chosen  distance and  velocity data
points  imply  a  central  mass   of  $4.6  \pm  0.7  \pm  0.3  \times
10^{11}\msun$  (the  second uncertainty  quantifies  the  effect of  a
$300\kms$ transverse  velocity).  A slightly more realistic  case is a
logarithmic potential  $\Psi = {{1}\over{2}}  v_c^2 \log( r_c^2  + r^2
)$, which  has a circular velocity  that asymptotes to  $v_c$ at large
radius  $r$.   Following \cite{dehnen},  we  adopt  a  core radius  of
$r_c=3\kpc$ for this model.  In this case,
$$v_c^2 = (v_2^2 - v_1^2) / \ln \Big( {{r_c^2 + r_1^2}\over{r_c^2 + r_2^2}}
\Big) ,$$
which gives a circular velocity of $v_c  = 162 \pm 15 \pm 8 \kms$.  We
also  investigated a simple  model in  which M31  is an  NFW potential
$\Psi= -G  M_s \log (1+r/r_s)  / r $,  where $M_s$ is the  mass within
$5.3 r_s$, and $r_s$ is  the scale radius. Assuming the characteristic
relation between halo mass and concentration in a $z=0$ LCDM cosmology
$c = 15.0 -  3.3 \log(M_{200}/10^{12}\msun h^{-1})$ \cite{bullock}, we
find $M_s  = 4.4 \pm  1.2 \pm 0.3 \times  10^{11}\msun$, ($M_{200}=8.0
\pm 2.1 \pm 0.5 \times 10^{11}\msun$, $r_{200}=194 \pm 16 \pm 4 \kpc$,
$v_{200}=136 \pm 11 \pm 3\kms$,  $r_s=13.1 \pm 1.4 \pm 0.4\kpc$).  For
comparison,   the  mass  inside   $125\kpc$  is   $7.6  \pm   1.2  \pm
0.3\times10^{11}\msun$ for  the logarithmic model and  $6.4\pm 1.3 \pm
0.3\times10^{11}\msun$ for the NFW model, respectively.  The projected
distance-velocity behaviour  of these three toy models  is compared to
the velocity  measurements in  Figure~1.  It is  no surprise  that the
Kepler potential  over-predicts the radial velocity close  to M31, but
both the simple logarithmic and  NFW halo models manage to approximate
the stream velocity profile well.

\begin{figure*}
\ifthenelse{\UseFigs=1}{
\includegraphics[angle=270,width=\hsize]{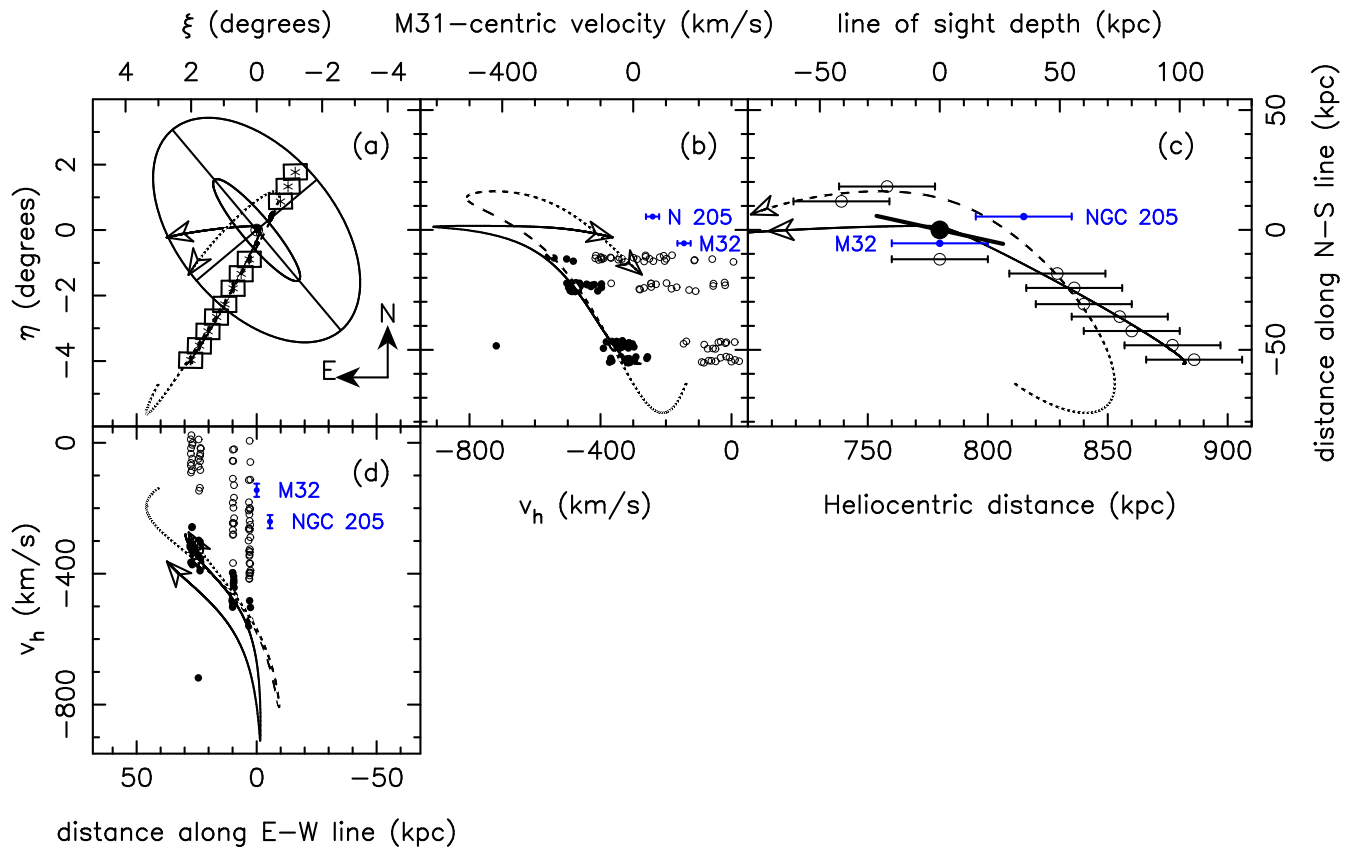}}{}
\caption{A multidimensional view of the two best-fitting orbits in the
more  realistic galaxy  model of  \citet{klypin02}. The  kinematics in
Figure~1, combined with the  distance information of Figure~2, require
that  the stream  orbit is  currently moving  in towards  M31, gaining
velocity from  Field 1 to Field  6 as the stream  approaches M31.  The
full line shows  the orbit projection on the  sky for the best-fitting
orbit, in  the best  fitting potential, when  only the data  in Fields
1--8 are taken into account; in this case the total mass of M31 inside
$125\kpc$ is $M_{125} =  7.5^{+2.5}_{-1.3} \times 10^{11} \msun$.  The
dashed  line however, shows  the best  fit orbit  in the  best fitting
potential when we also account for  the distance data in Fields 12 and
13.  Now the mass of the model is $M_{125} = 1.5\pm 0.1 \times 10^{12}
\msun$  inside  $125\kpc$.   In  both  cases the  curves  display  the
centre-of-mass orbits, integrated for $\pm 0.75\Gyr$ from the position
of the centre of Field 6; the arrowheads show the direction of motion.
Panel (a) shows the location on the sky of the two orbits superimposed
on the chart previously shown  in Figure~1; panels (b) and (d) present
projections   of  the  distance-velocity   relation  (symbols   as  in
Figure~1); and finally  panel (c) shows a sideways  view of the orbits
(as in Figure~2).}
\end{figure*}

To  constrain more  realistic models  of the  potential, we  adopted a
recent composite  galaxy model by \citet{klypin02},  and calculate the
potential due  to a  sum of disk,  bulge and halo  mass distributions.
The bulge in this model  has a mass of $1.9\times 10^{10}\msun$, while
the disk has  a mass of $7\times 10^{10}\msun$, and  a scale length of
$5.7\kpc$.   Fixing  the  disk  and bulge  with  the  \citet{klypin02}
parameters (with the further assumption  that the disk scale height is
$400\pc$,  as  suggested  by  \citealt{gould94}), we  investigate  the
spherical   halo   models  compatible   with   our  data.    Following
\citet{klypin02}, we  take into  account the adiabatic  contraction of
the halo due to the settling of baryons into the disk and bulge, which
alters the potential of the inner galaxy significantly.

We consider a halo of a given mass, with the appropriate concentration
\citep{bullock},  and  compress the  halo  using the  \citet{klypin02}
relations.   We then  calculate the  resulting potential  by multipole
expansion. Using an ``AMOEBA'' minimization algorithm \citep{press} we
launch  orbits  in the  potential,  iteratively  improving on  guessed
values for the initial position and velocity of a test particle on the
orbit. A $\chi^2$  goodness of fit statistic is  minimized between the
observed sky position, line of  sight depth and radial velocity of the
stars in our survey and the  values predicted by the orbit model. This
of  course  assumes  that the  stream  follows  the  orbit of  a  test
particle, an  assumption that  is reasonable if  its progenitor  was a
relatively  low-mass  dwarf  galaxy,  which  is  consistent  with  the
measured $11\pm 3\kms$ velocity dispersion in the stream.

We assume  that the orbit must  pass through the field  centres on the
sky to within $0.15^\circ$ ($2\kpc$); this corresponds to our estimate
of the uncertainty  in the measurement of the  location of the central
peak along  the stream.  To constrain  the distance of  the stream, we
take  the  measurements  of  \citet{mcconnachie} together  with  their
$20\kpc$  distance  uncertainties   (these  uncertainties  include  an
estimate of  the systematic error  in the distance  measurement).  The
fit is also  constrained with the measured radial  velocities; we take
all data-points within $100\kms$  of the straight-line fit in Figure~1
(those  that are  shown with  filled  circles), and  adopt the  fitted
velocity  dispersion of $11\kms$  as the  expected uncertainty  in the
velocity fit.

The  relative  likelihood of  halo  mass  models  can be  analyzed  by
comparing the likelihood of the orbit  fit as a function of halo mass.
However,  we find  that the  result  depends strongly  on whether  the
distance data  of Fields 12  and 13 are  included in the the  fit (the
stream RGB population was not detected in Field 14).

We first  investigate the consequences of rejecting  the distance data
in Fields  12 and  13.  In those  fields \citet{mcconnachie}  may have
detected the metal-rich  thick disk, or a warp in  the disk instead of
the actual stream itself. Without confirmation from radial velocities,
we cannot be certain that the  stream continues into that area of M31.
Fitting only the data  in Fields 1 to 8, we find  that the most likely
model  has a  total mass  inside $125\kpc$  of $M_{125}  =  7.5 \times
10^{11} \msun$;  the likelihood  drops by a  factor of  $e^{-1}$ (i.e.
$1\sigma$) for $M_{125}  = 1.0 \times 10^{12} \msun$.   We also reject
at the 99\%  confidence level a mass lower than  $M_{125} = 5.4 \times
10^{11} \msun$.  The most likely  orbit in this model galaxy potential
is shown as a full line  in the projections displayed in Figures~4 and
5, and  its circular velocity  curve is displayed  in the same  way in
Figure~6.   This orbit  manages  to fit  the  distance information  in
Fields 1--7 very well, but  it is peculiar in being exceedingly radial
(peri- to apocentre  ratio of 70). Indeed, the  orbit has a tangential
velocity in Field~1,  relative to M31, of $5\kms$,  that is, the orbit
has  only a  very small  non-radial velocity  component.  This  may be
partly responsible for the good agreement of the present mass estimate
with  those derived above  for the  three toy  models in  which radial
orbits were  imposed.  In the  present analysis we have  again assumed
that the tangential motion of  M31 is negligible.  However, the effect
of a $300\kms$  tangential velocity along the direction  of the Stream
alters the derived  mass inside $125\kpc$ by only  $0.5 \times 10^{11}
\msun$.

\begin{figure}
\ifthenelse{\UseFigs=1}{
\includegraphics[angle=270,width=\hsize]{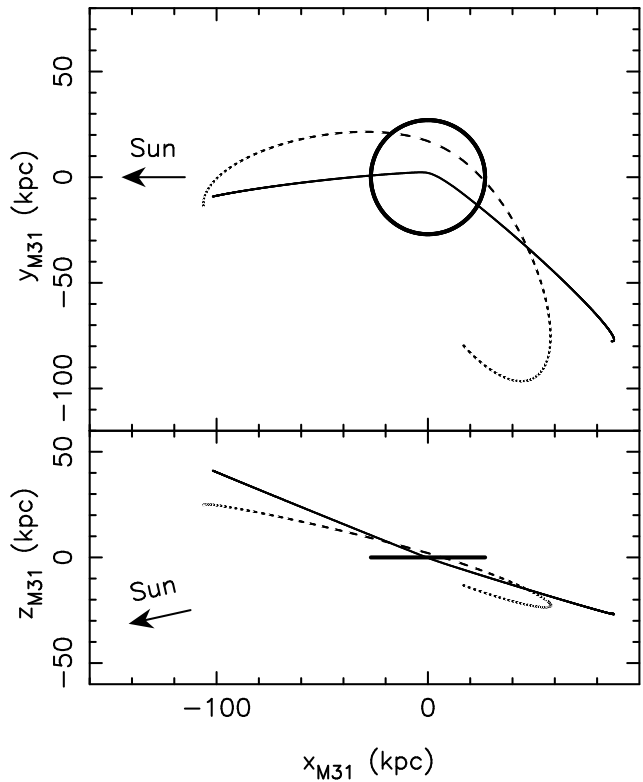}}{}
\caption{The   paths,   in  M31-centric   coordinates,   of  the   two
best-fitting orbits previously shown  in Figure~4.  In this coordinate
system,   the   Sun  is   located   at  $(x_{M31},y_{M31},z_{M31})   =
(-762,0,-169)\kpc$. The upper  panel shows a face-on view  of M31 (the
limits of  a $27\kpc$  radius disk is  indicted with a  thick circle),
while the lower panel is an edge-on view of the plane of M31.}
\end{figure}

If we decide  to also fit the  distance data in Fields 12  and 13, the
most likely model has a  much higher total mass inside $125\kpc$, with
$M_{125} = 1.5 \pm 0.1 \pm  0.25 \times 10^{12} \msun$ (again, the the
second  uncertainty  value reflects  a  tangential  motion  of M31  of
$300\kms$).  The  most likely orbit  in this potential model  is shown
with a  dashed line in Figures~4  and 5, and the  circular velocity of
the potential  is shown  with a dashed  line in Figure~6.   This orbit
does not manage  to fit the large line of sight  distances in Fields 1
and 2  as well as the  previous case (although the  discrepancy is not
statistically significant given the large error bars on these distance
data).  The  main shortcoming  is that the  circular velocity  of this
model at galactocentric  distances below $30\kpc$ is at  odds with the
data displayed in  Figure~6.  The circular velocity of  the halo alone
(without the bulge and disk) is higher than the data allow. This model
is therefore not acceptable.

\begin{figure}
\ifthenelse{\UseFigs=1}{
\includegraphics[angle=270,width=\hsize]{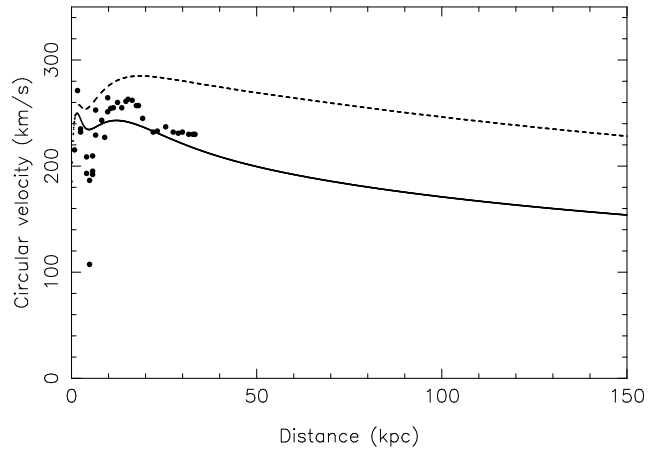}}{}
\caption{The   circular  velocity  curve   of  the   two  best-fitting
potentials. The high-mass model (dashed line), based on the assumption
that the stream  continues to Fields 12 and  13, greatly over-predicts
the observed rotation curve  between $20$--$30\kpc$.  The M31 rotation
curve data are  reproduced here with solid dots  (from the compilation
by \citealt{klypin02}).}
\end{figure}

\section{Discussion and conclusions}

In  this first  kinematic study  of the  giant stellar  stream  in the
Andromeda galaxy,  we have  been able to  measure the  radial velocity
gradient along  the stream from the outermost  field currently probed,
$125\kpc$ distant from the centre of M31, down to an inner field $\sim
20\kpc$ from the centre of  that galaxy.  Over this huge distance, the
(projected)   radial  velocity  changes   by  $245\kms$,   implying  a
de-projected  velocity  difference of  $\sim  280\kms$. This  velocity
gradient is  used to  obtain a zeroth  order analytic estimate  of the
mass of the  halo, which for simple halo-only galaxy  models such as a
logarithmic halo  or an NFW halo,  implies a mass  inside $125\kpc$ of
$M_{125} =  7.6 \pm 1.2 \times  10^{11} \msun$ and $M_{125}  = 6.4 \pm
1.3  \times   10^{11}  \msun$,   respectively.  In  both   cases,  the
uncertainty associated  with a possible tangential velocity  of M31 of
$300\kms$, is $\sim 5$\%.

We also  investigate more realistic solutions, allowing  the stream to
have a non-radial orbit, and taking  a galaxy model that is the sum of
a disk,  bulge and dark halo  \citep{klypin02}. The dark  halo of this
model is a  perturbation on a spherical NFW model,  to account for the
adiabatic contraction  of the dark  matter as the  baryonic components
form.  If we disregard the distance data in Fields 12 and 13, since we
cannot be  certain that  the stream is  present in those  regions, the
most likely mass of this  galaxy model is $M_{125} = 7.5^{+2.5}_{-1.3}
\times  10^{11}\msun$, with  a lower  limit of  $M_{125} =  5.4 \times
10^{11} \msun$  (at 99\% confidence). This result  is fully consistent
with the zeroth order analytic estimates discussed above, and suggests
that the derived mass is not very sensitive to the adopted mass model.
Furthermore, there is a reasonable agreement of the resulting rotation
curve  with the  kinematics of  previously-observed disk  tracers (see
Figure~6),  despite  the fact  that  the model  was  only  fit to  the
kinematics of the  stream.  This agreement is due in  part to the fact
that we  have taken previously-fitted  models for the disk  and bulge,
but the halo contribution to the total rotation curve dominates beyond
$\sim 20\kpc$, and it is at  these large distances that we have fitted
the model to the stream  data.  This confers further confidence on the
derived mass  model. The uncertainty on  the mass estimate  due to the
possible  tangential  velocity  of  M31  is  likely  not  very  large,
approximately 7\% for a tangential  velocity of $300\kms$.  One of the
main uncertainties  in the present  analysis is the flattening  of the
halo, which  we have not  explored, as the  current data set  does not
provide  sufficient  constraints.    Future  studies,  fitting  N-body
simulations  to a  larger kinematic  sample  of stream  stars, can  be
expected  to improve  the mass  estimate and  also constrain  the halo
shape.

\begin{figure}
\ifthenelse{\UseFigs=1}{
\includegraphics[angle=0,width=\hsize]{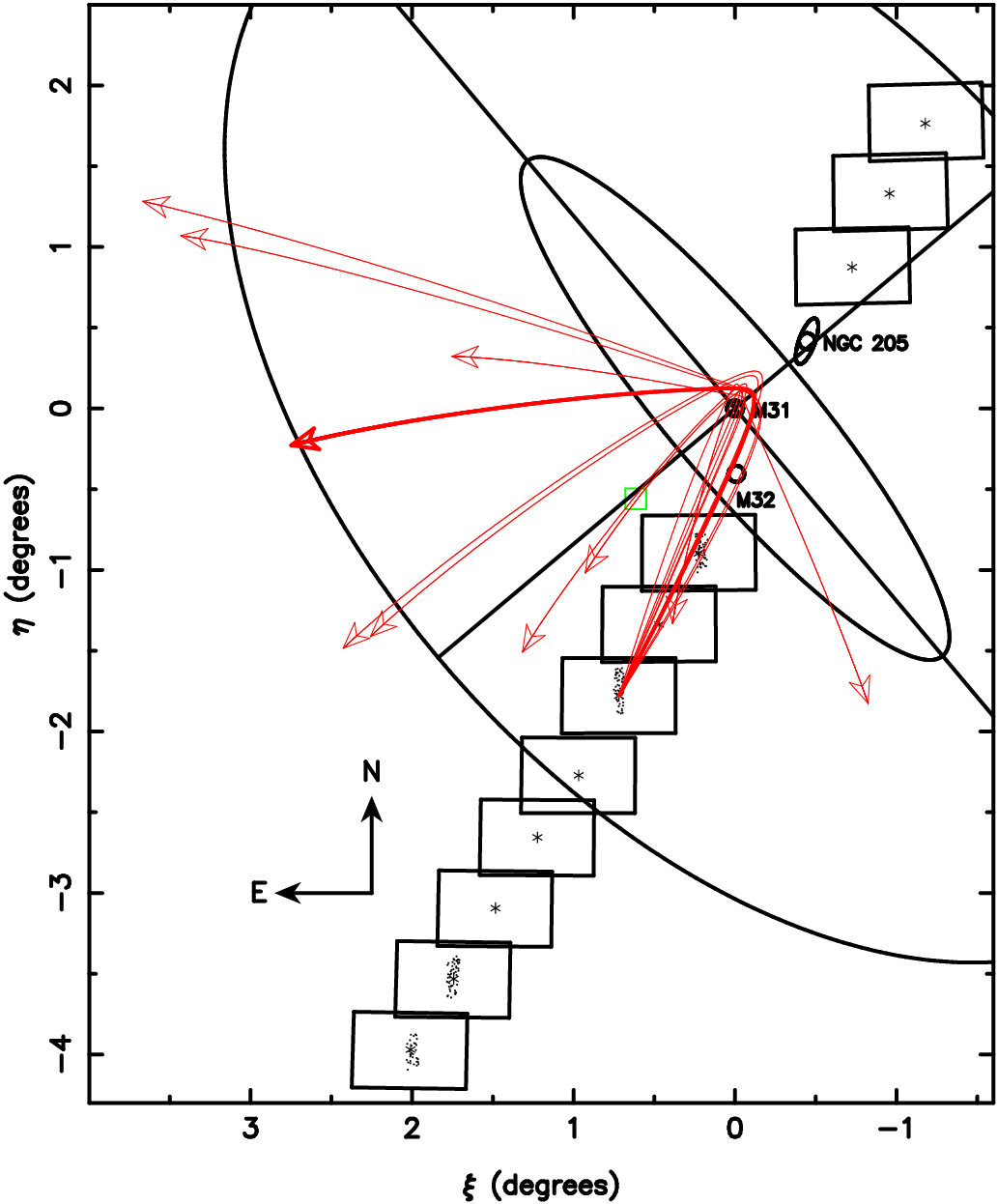}}{}
\caption{The spreading of orbits after a close encounter with M31. The
thick-line orbit reproduces the best-fit orbit in the realistic galaxy
potential previously shown in  Figure~4, integrated from the centre of
Field~6.  The  thin-lines show similar orbits, starting  from the same
spatial  position, but with  velocities perturbed  by a  random offset
drawn  from  a Gaussian  distribution  of  dispersion $11\kms$.   This
fanning-out of stars  on nearby orbits will lead  to the disappearance
of  the  Andromeda  Stream,  and   implies  that  it  is  a  transient
phenomenon. The small  square on the minor axis  shows the position of
the \citet{brown} ACS field (not  drawn to scale!). All other markings
are as in Figure~1.}
\end{figure}

It is  only recently that measurements  of the mass of  M31 beyond the
edge of its gaseous disk have been possible. \citet{courteau} analysed
the  velocities of  7 Andromeda  satellites, finding  a total  mass of
$13.3\pm 1.8 \times 10^{11} \msun$. In contrast, using a larger sample
of 10 satellite galaxies, 17 globular clusters and 9 planetary nebulae
as test  particles, \citet{evans00a} found that the  most likely total
mass  of M31 is  $12.3^{+18}_{-6} \times  10^{11}\msun$, approximately
half   of  their   Milky  Way   estimate  of   $19^{+36}_{-17}  \times
10^{11}\msun$.  With improved radial  velocities of the M31 satellites
they were later able to reduce their M31 mass uncertainties, finding a
value   of  the   total   mass  of   $\sim  7.0^{10.5}_{-3.5}   \times
10^{11}\msun$  \citep{evans00b}, which  is fully  consistent  with our
result of  $M_{125} = 7.5^{+2.5}_{-1.3}  \times 10^{11}\msun$ (defined
within a radius of $125\kpc$).  However, a definitive statement of the
relative  masses  of  M31  and   the  Milky  Way  awaits  an  improved
measurement for the Milky Way.

The best  fit orbit in the best  fit potential is prograde  and in the
region where it  is currently observed, it lies close  to the plane of
M31.   Thus it  appears  that the  orbit  of the  Andromeda stream  is
peculiar  in  being  extremely  radial,  passing  very  close  (within
$2\kpc$) of  the centre  of M31.  This  requires very  special initial
conditions.  The  stream stars, which are  spatially narrowly confined
in Fields~1 to 8, will  diverge dramatically upon passing close to M31
to form  a low-density fan-like  structure, since orbits  that deviate
only slightly  from the  orbit displayed in  Figure~4 on  the plunging
part  of their  course  will  take very  different  paths after  being
accelerated around the centre of M31, as demonstrated in Figure~7. The
fanning-out of the stream is  likely to confuse efforts to measure the
metallicity  and  age  of  the  M31 halo;  for  instance,  the  recent
discovery  by \citet{brown}  of a  young  halo component  in M31  from
main-sequence fitting  of an  extremely deep ACS  field may be  due to
stream contamination (see Figure~7).

Thus  the stream  may be  in the  process of  vanishing as  a coherent
structure, providing a supply of metal-rich stars into the halo.  This
also suggests  that the  stream was removed  from its  progenitor less
than an orbital period ago (the pericentre to pericentre period of the
continuous-line orbit in Figure~4 is $1.8\Gyr$), as we would otherwise
not observe  the structure as a  stream.  The ephemeral  nature of the
stream  implies that  the progenitor  must have  survived  until $\sim
1.8\Gyr$ ago. As we discuss further below, the progenitor was probably
of  low mass, implying  that the  rate of  decay of  its orbit  due to
dynamical friction was  slow, so it followed (or  continues to follow)
an orbit close to the current orbit of the stream.  However, any dwarf
galaxy on the derived orbit  must have experienced extreme tides as it
repeatedly passed close to the centre  of M31.  One option is that the
progenitor was a very dense  dwarf galaxy that was sufficiently robust
to survive  the huge tides.  This  brings to mind M32  as a candidate,
though  detailed  numerical  modeling  is  required  to  examine  this
possibility.   The alternative option  is that  the progenitor  of the
stream deflected off another halo object, sending it plunging into the
current orbit, analogous to  the suggestion by \citet{zhao98} to explain
the longevity of the Galactic satellite Sagittarius.

However, the connection with M32 presents some difficulties.  Although
M32  appears  to  reside  in  the stream,  its  velocity  is  markedly
different. For M32  to be associated with the  stream would require it
to be at a different phase  in the orbit (either lagging or trailing).
Furthermore, the low velocity dispersion of the stream would appear to
preclude  M32 as  its  progenitor (which  has  $\sigma_v \sim  50\kms$
outside of  the nucleus, \citealt{vandermarel94}),  though this cannot
be  confirmed  without  a  detailed  dynamical study.   The  case  for
association of NGC~205 with the stream also appears weaker given these
kinematic measurements, since the best fit orbit does not overlap with
it in phase-space.  Future studies  may allow us to examine this issue
in more  detail by  following the stream  beyond the  region currently
probed with kinematics.  The  velocities of the stream stars presented
here also  shed light on the  recent identification of  a possible new
companion to M31, And VIII \citep{morrison}.  The radial velocities of
those  planetary  nebulae,  which   are  located  close  to  M32  (see
Figure~1),  appear  to  have  radial  velocities  consistent  with  an
extrapolation of the Stream, as they lie close to the straight line in
the right-hand panel of Figure~1.  This would suggest that And VIII is
most likely  part of the Andromeda  Stream, situated in  the region of
highest  over-density reported  by \citet{ibata01b}.   However,  it is
interesting to note that the radial velocities of the \citet{morrison}
planetary nebula sample tend to increase towards the North (showing an
apparent  positive  gradient in  the  right-hand  panel of  Figure~1),
whereas  the Stream  stars  have a  negative  velocity gradient.   The
connection between the two  structures therefore merits to be examined
more   carefully.   During   the  refereeing   process,  a   study  by
\citet{merrett}  was presented which  also investigates  the planetary
nebulae around  M31.  However, though  their survey has  also detected
PNe in  a region at the base  of the stream (near  our Field~8), their
interpretation  is  inconsistent  with  the kinematics  of  the  stars
reported here.   The direction of motion  of the orbit  they derive is
opposite to ours,  and the path of their  orbit, which intercepts many
M31-disk  PNe is  substantially different  to the  orbit that  we have
fitted.  It  is possible  that their finding  reveals the  presence of
another kinematic structure in Andromeda.

The  majority of the  stream stars  that were  surveyed have  a narrow
velocity  dispersion  of  $11\pm  3\kms$, though  slightly  skewed  to
positive velocities.  The  fact that our line of  sight looks down the
stream (see Figure~2),  so that we probably see stars  over a range of
distance along the line of sight, and hence at different phases in the
orbit,  will tend to  render the  observed velocity  dispersion higher
than  the  intrinsic velocity  dispersion.   This  indicates that  the
progenitor was  most likely  a low mass  dwarf galaxy.  The  Milky Way
satellite  Sagittarius, which  has a  velocity dispersion  of $11\kms$
\citep{ibata97}, also has a gigantic stellar stream, but with a larger
velocity dispersion of $20\kms$  \citep{yanny}. However, it is unclear
at  present whether  the  lower velocity  dispersion  measured in  the
Andromeda stream  compared to the Sagittarius stream  implies that its
progenitor was of lower mass than Sagittarius, or not.

\section*{Acknowledgments}

\medskip

RI would like  to thank A. Klypin and H.-S. Zhao  for kindly giving us
their  compilation of  M31  disk kinematics,  and  for explaining  the
details  of their  Andromeda galaxy  model. The  anonymous  referee is
thanked for comments that improved the paper. The research of AMNF has
been supported by  a Marie Curie Fellowship of  the European Community
under contract number HPMF-CT-2002-01758.

\newcommand{\mnras}{MNRAS}
\newcommand{\nat}{Nature}
\newcommand{\araa}{ARAA}
\newcommand{\aj}{AJ}
\newcommand{\apj}{ApJ}
\newcommand{\apjl}{ApJ}
\newcommand{\apjs}{ApJSupp}
\newcommand{\aap}{A\&A}
\newcommand{\aaps}{A\&ASupp}
\newcommand{\pasp}{PASP}


\begin{thebibliography}{}
%
\bibitem[Brown et al.(2003)]{brown}
        Brown, T., Ferguson, H., Smith, E., Kimble, R., Sweigart, A., 
        Renzini, A., Rich, M., VandenBerg, Don A., 2003, \apj\ 592, 17L
%
\bibitem[Bullock et al.(2001)]{bullock}
        Bullock, J., Kolatt, T., Sigad, Y., Somerville, R., Kravtsov,
        A., Klypin, A., Primack, J., Dekel, A., 2001, MNRAS 321, 559
%
\bibitem[Courteau \& van den Bergh(1999)]{courteau}
        Courteau, S., van den Bergh, S., 1999, \aj\ 118, 337
%
\bibitem[Dehnen \& Binney(1998)]{dehnen}
	Dehnen, W. \& Binney, J., 1998, \mnras\ 294, 429
%
\bibitem[Einasto \& Lynden-Bell(1982)]{einasto}
        Einasto, J., Lynden-Bell, D., 1982, \mnras\ 199, 67
%
\bibitem[Evans \& Wilkinson(2000)]{evans00a}
        Evans, N., Wilkinson, M., 2000, \mnras\ 316, 929
%
\bibitem[Evans et al.(2000)]{evans00b}
        Evans, N., Wilkinson, M., Guhathakurta, P., Grebel, E., Vogt, S., 2000, \apj\ 540, 9L
%
\bibitem[Ferguson et al.(2002)]{ferguson02}
        Ferguson, A., Irwin, M., Ibata, R., Lewis, G., Tanvir,
        N., 2002, \aj\ 124, 1452
%
\bibitem[Gould(1994)]{gould94}
        Gould, A., 1994, \apj\ 435, 573
%
\bibitem[Helmi \& White(1999)]{helmi99}
	Helmi, A., White, S., 1999, \mnras\ 307, 495
%
\bibitem[Ibata et al.(1997)]{ibata97}
	Ibata, R., Wyse, R., Gilmore, G., Irwin, M. \& Suntzeff, N., 1997,
	\aj\ 113, 634
%
\bibitem[Ibata et al.(2001a)]{ibata01a}
	Ibata R., Lewis G., Irwin M., Totten E. \& Quinn T., 2001, \apj\ 551, 294
%
\bibitem[Ibata et al.(2001b)]{ibata01b}
        Ibata, R., Irwin, M., Lewis, G., Ferguson, A., Tanvir,
        N., 2001, Nature 412, 49
%
\bibitem[Ibata et al.(2002)]{ibata02}
	Ibata R., Lewis G., Irwin M. \& Cambr\'esy L., 2002, \mnras\ 332, 921
%
\bibitem[Johnston et al.(1999)]{johnston99}
        Johnston, K., Zhao, H.-S., Spergel, D., Hernquist, L., 1999, \apj\ 512, 109L
%
\bibitem[Johnston, Sackett \& Bullock(2001)]{johnston01}
        Johnston, K., Sackett, P., Bullock, J., 2001, \apj\ 557, 137
%
\bibitem[Kahn \& Woltjer(1959)]{kahn}
        Kahn, F., Woltjer L., 1959, \apj\ 130, 705
%
\bibitem[Klypin, Zhao \& Somerville(2002)]{klypin02}
        Klypin, A., Zhao, H.-S., Somerville, R., 2002, \apj\ 573, 597
%
%
\bibitem[Lynden-Bell \& Lin(1977)]{DLB77}
        Lynden-Bell, D., Lin, D., 1977, \mnras\ 181, 37
%
\bibitem[Majewski et al.(2003)]{majewski}
        Majewski S., Skrutskie M., Weinberg M. \& Ostheimer J., 2003, {\it astro-ph/0304198}
%
\bibitem[Malin \& Hadley(1997)]{malin}
        Malin, D., Hadley, B., 1997, PASA 14, 52
%
\bibitem[van der Marel et al.(1994)]{vandermarel94}
        van der Marel, R., Rix, H.-W., Carter, D., Franx, M., White,
        S., de Zeeuw, T., 1994, \mnras\ 268, 521
%
%
\bibitem[McConnachie et al.(2003)]{mcconnachie}
	McConnachie, A., {Irwin}, M., {Ibata}, R., Ferguson, A.,
	{Lewis}, G., Tanvir, N., 2003, \mnras\ 343, 1335
%
\bibitem[Merrett et al.(2003)]{merrett}
        Merrett, H., Kuijken, K.,  Merrifield, M., Romanowsky, A., Douglas,
        N., Napolitano,  N., Arnaboldi,  M., Capaccioli, M.,  Freeman, K.,
        Gerhard, O.,  Evans, N., Wilkinson, M., Halliday,  C., Bridges, T.,
        Carter, D., 2003, astro-ph/0311090
%
\bibitem[Morrison et al.(2003)]{morrison}
        Morrison, H., Harding, P., Hurley-Keller, D., Jacoby, G., 2003,
        \apj\ 596, 183L
%
\bibitem[Navarro, Frenk \& White(1997)]{navarro}
        Navarro, J., Frenk, C., White, S., 1997, \apj\ 490, 493
%
\bibitem[Pohlen et al.(2003)]{pohlen}
        Pohlen, M., Martinez-Delgado, D., Majewski, S., Palma, C.,
        Prada, F., Balcells, M., 2003, in 
        "Satellites and Tidal Streams" ASP conference, La Palma, Canary
        Islands, 26-30 May 2003, eds, F. Prada, D. Martinez-Delgado, T. Mahoney
%
\bibitem[Press et al.(1992)]{press}
	Press, W., Flannery, B., Teukolsky, S. and Vetterling, W., 1992
	`Numerical Recipes' (Cambridge Univ Press, Cambridge)
%
\bibitem[Reitzel \& Guhathakurta(2002)]{reitzel}
        Reitzel, D., Guhathakurta, P., 2002, \aj\ 124, 234
%
\bibitem[Stanek \& Garnavich(1998)]{stanek}
        Stanek, K., Garnavich, P., 1998, \apj\ 503, 131L
%
\bibitem[de Vaucouleurs et al.(1991)]{devaucouleurs}
        de Vaucouleurs, G., de Vaucouleurs, A., Corwin, H., Buta, R.,
        Paturel, G., Fouque, P., 1991, "Third Reference Catalogue of
        Bright Galaxies", Springer-Verlag, Berlin Heidelberg New York
%
\bibitem[Yanny et al.(2003)]{yanny}
        Yanny, B., et al., 2003, \apj\ 588, 824
%
\bibitem[Zhao(1998)]{zhao98} 
        Zhao, H.-S., 1998, \apj, 500, L149
%
\bibitem[Zhao et al.(1999)]{zhao99} 
        Zhao, H.-S., Johnston, K., Henquist, L., Spergel, D., 1998,
        \aap\ 348, 49
%
\end{thebibliography}
\end{document}